\documentstyle[preprint,aps,prc,floats,psfig,tighten,epsfig,axodraw,epic]{revtex}
\draft
\renewcommand{\thefootnote}{\arabic{footnote}}
\def\gev{\rm GeV}
\def\fbi{\rm fb^{-1}}
\def\epem{e^+e^-}
\def\mh{m_h^{}}
\def\lsim{\mathrel{\raise.3ex\hbox{$<$\kern-.75em\lower1ex\hbox{$\sim$}}}}
\def\gsim{\mathrel{\raise.3ex\hbox{$>$\kern-.75em\lower1ex\hbox{$\sim$}}}}
\def\Re{{\cal R}\!e}
\def\Im{{\cal I}\!m}

\begin{document}

\draft
\renewcommand{\thefootnote}{\arabic{footnote}}

\preprint{
\vbox{\hbox{\bf MADPH-00-1201}
      \hbox{\bf hep-ph/0011271}
      \hbox{November, 2000}}}

\title{\bf CP-violating $ZZh$ Coupling at $\epem$ Linear Colliders}
\author{T. Han\footnote{than@egret.physics.wisc.edu}
and J. Jiang\footnote{jiang@pheno.physics.wisc.edu}}
\address{Department of Physics, University of Wisconsin,\\ 
1150 University Avenue, Madison, WI 53706, USA}

\maketitle

%\vskip 0.5cm

\begin{abstract}
We study the general Higgs-weak boson coupling with 
CP-violation via the process $e^+e^-\to f\bar f h$.
We categorize the signal channels by sub-processes
$Zh$ production
and $ZZ$ fusion and construct four CP asymmetries 
by exploiting polarized $\epem$ beams. We find
complementarity among the sub-processes and the
asymmetries to probe the
real and imaginary parts of the CP-violating form
factor. Certain asymmetries with unpolarized beams 
can retain significant sensitivity to the coupling. 
We conclude that at a linear collider with high 
luminosity, the CP-odd $ZZh$ coupling may be sensitively 
probed via measurements of the asymmetries.
\end{abstract}
\newpage

\section{Introduction}

Searching for a Higgs boson has been a major motivation
for many current and future collider experiments, since
Higgs bosons encode the underlying physics of mass generation.
In the minimal Standard Model (SM), there is only one 
CP-even scalar. In the two-Higgs doublet model
or the supersymmetric extension of the SM, there are
two CP-even states, one CP-odd state, plus a pair
of charged Higgs bosons. 
The couplings of Higgs bosons to electroweak 
gauge bosons are particularly important since they faithfully 
represent the nature of the electroweak gauge symmetry breaking. 
Determining the detailed properties of the Higgs boson couplings 
will be of fundamental importance to fully construct the 
theoretical framework of the electroweak sector. 

\begin{figure}[tbh]
%\centerline{\psfig{file=/u/than/temp/Zh.eps,width=2.8in}}
%\centerline{\psfig{file=Zh-color.eps,width=2.8in}}
%file=Zh-color.eps,width=2.8in}}
\begin{center}
\begin{picture}(160,100)(0,0)
\dashline{5}(80,50)(140,50)
\Photon(30,10)(80,50){3}{6}
\Photon(30,90)(80,50){4}{6}
\Text(10,10)[]{$Z^{\nu}(k_2)$}
\Text(10,90)[]{$Z^{\mu}(k_1)$}
\Text(105,35)[]{$\Gamma^{\mu\nu}(k_1,k_2)$}
\Text(150,50)[]{$h$}
\end{picture}
\end{center}
\smallskip
\caption{Vertex function for $ZZh$ coupling.
\label{vert}}
\end{figure}
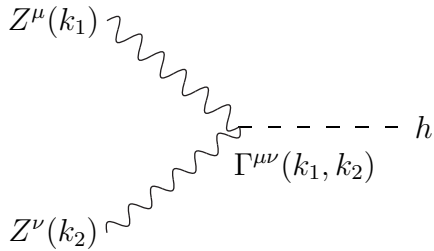

The most general interaction vertex for a generic 
Higgs boson ($h$) and a pair of $Z$ bosons,
$Z^\mu (k_1)\ Z^\nu (k_2)\ h$,
can be expressed by the following Lorentz structure
\begin{equation}
\Gamma^{\mu\nu}(k_1,k_2) = i {2\over v}\ h \ 
[a\  M_Z^2 g^{\mu\nu} + b\ (k_1^\mu k_2^\nu -k_1\cdot k_2 g^{\mu\nu}) +
\tilde b\ \epsilon^{\mu\nu\rho\sigma}k_{1\rho} k_{2\sigma}],
\label{vertx}
\end{equation}
where $v=(\sqrt 2 G_F)^{-1/2}$ is the vacuum expectation value 
of the Higgs
field, and the $Z$ boson four-momenta are both incoming, as depicted
in Fig.~{\ref{vert}. The $a$ and $b$ terms are CP-even and the
$\tilde b$ term is CP-odd. Thus, the simultaneous existence of 
terms $a$ (or $b$) and $\tilde b$ would indicate CP violation
for the $ZZh$ coupling \cite{keung,parity,eezh}. 
We note that in the SM at tree level, 
$a=1$ and $b=\tilde b=0$. In supersymmetric theories 
with CP-violating soft SUSY breaking terms \cite{carlos}, 
these CP-violating interactions may be generated
by loop diagrams. More generally, the parameters can be 
momentum-dependent form factors and of
complex values to account for the dispersive [$\Re(\tilde b$)] 
and absorptive [$\Im(\tilde b$)] effects from 
radiative corrections. Alternatively, in terms of an effective 
Lagrangian, the $b$ term
can be from gauge invariant dimension-6 operators \cite{dieter},
and the $\tilde b$ term can be constructed similarly with CP-odd
operators involving the dual field tensors. 
Dimensional analysis implies that the parameters 
$b$ and $\tilde b$ may naturally 
be of the order of $(v/\Lambda)^2$ where 
$\Lambda$ is the scale at which the physics responsible
for the electroweak symmetry breaking sets in, presumably
$\Lambda\lsim 4\pi v$. The CP-odd coefficient $\tilde b$
is of course very much model-dependent.

Possible CP-violation effects via Higgs-gauge boson 
couplings have recently drawn a lot of attention
in the literature. In Ref.~\cite{keung}, CP-odd observables
in decays $h\to ZZ, W^+W^-$ and $t\bar t$ were constructed.
It was discussed extensively how to explore the Higgs properties
via the process $\epem\to Zh$ \cite{parity,eezh} at future linear 
colliders. The polarized photon-photon collisions for 
$\gamma\gamma \to h$ \cite{gmgm} and the electron-electron 
scattering process $e^-e^-\to e^-e^- h$ \cite{emem}
were also considered to extract the CP-violating couplings.
There has also been considerable amount of work for 
investigation of CP-violating Higgs boson 
interactions with fermions at future $\epem$ 
colliders \cite{jack}. 

In this paper, we study the CP-violating coupling of
$ZZh$ at future $\epem$ linear colliders.
In Section II, we set out the general
consideration, identifying the $Zh$ production and
$ZZ$ fusion signals and exploring the generic
CP-odd variables by exploiting the polarized beams.
Given specific kinematics of the signal processes
under investigation, we construct four CP asymmetries
in Section III. We find important complementarity
among the sub-processes and 
the asymmetries in probing different aspects
of the CP-odd coupling, namely the real (dispersive)
and imaginary (absorptive) parts of $\tilde b$. 
We also examine to what 
extent this coupling can be experimentally probed
via measurements of the CP asymmetries, with and without
beam polarization. 
We present some general discussions of our analyses and
summarize our results in Section IV.

\section{General consideration}

We concentrate on the scenario with a light
Higgs boson below the $W$-pair threshold. The Higgs-weak boson
coupling will be studied mainly via Higgs production, rather
than its decay. We focus on the Higgs boson production associated 
with a fermion pair in the final state
\begin{equation}
e^-(p_1)\ e^+(p_2)\to f(q_1)\ \bar f(q_2)\ h(q_3).
\label{eeh}
\end{equation}
The Higgs boson signal may be best identified by examining
the recoil mass variable
\begin{equation}
m_{rec}^2=(p_1+p_2 - q_1 - q_2)^2
=s+m_{ff}^2-2{\sqrt s}(E_{f}+E_{\bar f}),
\label{rec}
\end{equation}
where $m_{ff}$ is the $f\bar f$ invariant mass and $E_f\ (E_{\bar f})$
is the fermion (anti-fermion) energy in the c.~m.~frame. This recoil 
mass variable will yield a peak for the signal at the Higgs mass
$\mh$, independent of the Higgs decay. This provides a
model-independent identification for the Higgs signal.
For this purpose, we will accept only 
\begin{equation}
f=e^-,\ \mu^-\quad {\rm and}\quad u,\ d,\ s
\label{fermions}
\end{equation}
to assure good energy determination for the final state
leptons and light quark jets. 
Whenever appropriate, we adopt energy smearing
according to a Gaussian distribution as
\begin{eqnarray}
\label{mree}
{\Delta E\over E}&=&{12\%\over \sqrt E} \oplus 1\%\qquad{\rm for\ leptons}\\
&=&{45\%\over \sqrt E} \oplus 2\%\qquad{\rm for\ quarks} .
\end{eqnarray}
In realistic experimentation, the charged tracking information 
may also be used to help improve the momentum determination.

As an illustration, the recoil mass spectrum for an $\epem$ final state
is shown in Fig.~\ref{mee} by the dashed curve. The width
of the peak in $m_{rec}$ spectrum is determined by the
energy resolution of the detector as simulated with Eq.~(\ref{mree}).
We have also required the final state fermions to be within the
detector coverage, assumed to be
\begin{equation}
|\cos\theta_f| < \cos10^\circ
\end{equation}
with respect to the beam hole. 

\begin{figure}[tb]
%\centerline{\psfig{file=/u/than/temp/fig2_mee.ps,width=3.6in}}
\centerline{\psfig{file=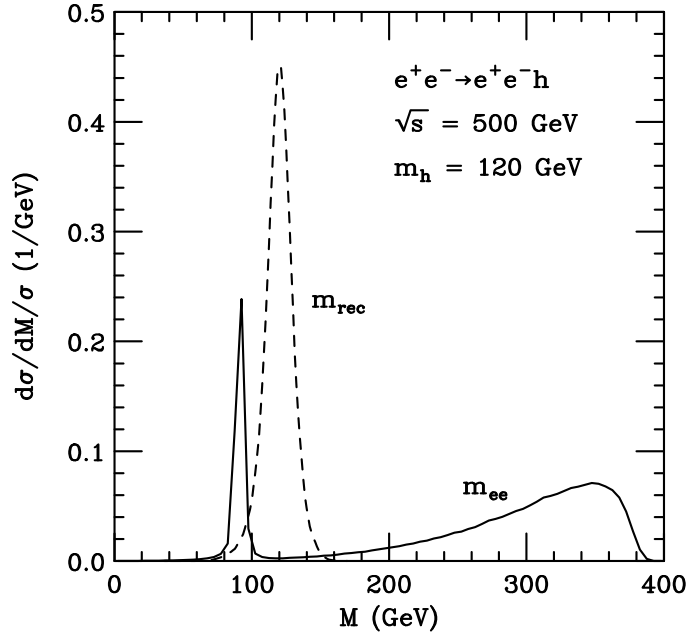,width=3.6in}}
\bigskip
\caption{Normalized mass distributions for $\epem 
\to \epem h$ at $\sqrt s=500\ \gev$ with
$\mh=120\ \gev$. The dashed curve is for the
recoil mass in Eq.~(\ref{rec}), and the solid is for the invariant
mass $m_{ee}$.}
\label{mee}
\end{figure}

\subsection{$Zh$ production versus $ZZ$ fusion}

The signal channel Eq.~(\ref{eeh}) can be approximately divided 
into two sub-processes
\begin{eqnarray}
\label{zh}
&& \epem \to Zh\qquad (Zh\ {\rm production})~,\\
\label{zz}
&& Z^*Z^* \to h\qquad (ZZ\ {\rm fusion}).
\end{eqnarray}
Eq.~(\ref{zh}) yields light fermion states of all flavors from $Z$ decay;
while Eq.~(\ref{zz}) always has an $\epem$ pair in the final state.
These two sub-processes can be effectively distinguished by
identifying the final state fermions. Even for the final state
of $\epem$, one can separate them 
by examining the mass spectrum $m_{ee}$. This is illustrated
in Fig.~\ref{mee} for the $\epem$ final state by the solid curve. 
The sharp peak at
$M_Z$ indicates the contribution from the decay $Z\to \epem$,
while the continuum spectrum at higher mass values is from
$ZZ$ fusion sub-process. In our analysis, we have
included both contributions coherently. However, when 
necessary, we separate out the $ZZ$ fusion contribution by requiring
\begin{equation}
m_{ee}>100\ \gev.
\label{cut0}
\end{equation}

\begin{figure}[tb]
%\centerline{\psfig{file=/u/than/temp/fig1_tot.ps,width=5.6in}}
\centerline{\psfig{file=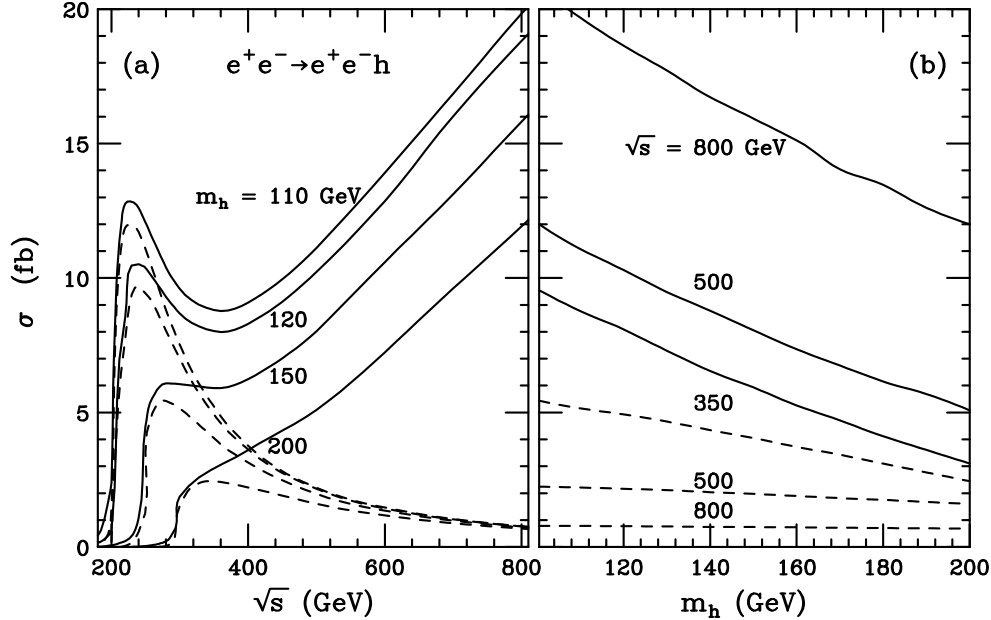,width=5.2in}}
\bigskip
\caption{Total cross sections for $\epem \to \epem h$ in fb
(a) versus $\sqrt s$ for representative values of $\mh$, and
(b) versus $\mh$ for representative values of $\sqrt s$.
The dashed curves are for $\epem \to Zh \to \epem h$
only. No kinematical cuts are imposed.}
\label{tot}
\end{figure}

The $Zh$ associated production is the leading channel for Higgs
boson searches at $\epem$ colliders. $ZZ$ fusion, on the other
hand, is often thought to be much smaller due to the small
$Z\bar e e$ vector coupling and low radiation rate of $Z$
bosons off $e^\pm$ beams. However, the rate of the fusion 
process increases with c.~m.~energy logarithmically
like $\ln^2(s/M_Z^2)$, and it is also 
more important for higher Higgs masses. The $ZZ$ fusion 
process naturally leads to a pair of electrons in the
final state, which is desirable when the charge information of
the final state is needed. 
Moreover, due to the helicity conservation at high energies,
the $Zh$ production has only helicity combinations for
the initial $\epem$ of $(+-)$ and $(-+)$; while the $ZZ$ fusion
has $(--)$ and $(++)$ in addition, where $-\ (+)$ refers
to the left (right) handed helicity. These additional helicity
amplitudes may provide further information regarding the CP test,
as we will see in the later analysis.

Figure \ref{tot} presents the total cross sections for 
$\epem \to \epem h$ to demonstrate
the comparison between the $Zh$ and $ZZ$ fusion processes.
Figure~\ref{tot}(a) gives cross sections versus
$\sqrt s$ for $\mh=110-200\ \gev$, and (b)
versus $\mh$ for $\sqrt s=350-800\ \gev$. The solid curves
are for the total SM rate including all contributions
coherently, and the dashed curves are with a real
$Z$ decay for $Zh\to \epem h$. We see that at 
$\sqrt s=500\ \gev$ and $\mh=120\ \gev$, 
$\sigma(ZZ\to h)\approx 10$ fb$>2\sigma(Zh\to \epem h,\ \mu^+\mu^- h)$. 
At $\sqrt s=800\ \gev$ and $\mh=120\ \gev$, the fusion 
cross section becomes about an order of magnitude higher
than that of $Zh\to \epem h,\ \mu^+\mu^- h$. Clearly, at
a linear collider above the $Zh$ threshold, the $ZZ$ fusion
process is increasingly more important in studying the Higgs 
properties \cite{ron}.

\subsection{CP property}

To unambiguously identify the effect of CP violation, 
one needs to construct a ``CP-odd variable'', 
whose expectation value vanishes if CP is conserved \cite{german}.
We begin our analysis by examining the CP-transformation
property. First of all, we note that the initial state 
of Eq.~(\ref{eeh}) can be made a CP eigenstate, given
the CP-transformation relation
\begin{equation}
e^-( \sigma_1,\vec p)\ e^+( \sigma_2,-\vec p) \Rightarrow 
e^-(-\sigma_2,\vec p)\ e^+(-\sigma_1,-\vec p),
\label{cp}
\end{equation}
where $\sigma_i$ is the fermion helicity.
Now consider a helicity matrix element
${\cal M}_{\sigma_1\sigma_2}(\vec q_1,\vec q_2)$
where $\sigma_1\ (\sigma_2)$ denotes the helicity of the
initial state electron (positron), which coincides with the
longitudinal beam polarization; $\vec q_1\ (\vec q_2)$ denotes 
the momentum of the final state fermion (anti-fermion).
It is easy to show that under CP transformation,
\begin{eqnarray}
\label{pm}
&&{\cal M}_{-+}( {\vec q_1}, {\vec q_2}) \Rightarrow
  {\cal M}_{-+}(-{\vec q_2},-{\vec q_1}),\\
&&{\cal M}_{--}({\vec q_1},{\vec q_2}) \Rightarrow
  {\cal M}_{++}(-{\vec q_2},{-\vec q_1}),
\label{ppmm}
\end{eqnarray}
and ${\cal M}_{+-}$, ${\cal M}_{++}$ transform similarly.
If CP is conserved in the reaction, Relations (\ref{pm}) 
and (\ref{ppmm}) take equal signs. 
These relations precisely categorize two typical classes of CP test:

\vskip 0.2cm
\noindent
\underline{\it CP eigen-process:} Under CP, 
${\cal M}_{-+}$ (or ${\cal M}_{+-}$) is invariant 
if CP is conserved. 
One can thus construct CP-odd {\it kinematical variables}
to test the CP property of the theory. We can construct a
``forward-backward'' asymmetry 
\begin{equation}
{\cal A}^{FB}=\sigma^F-\sigma^B =
\int_{0}^{1}{  d\sigma\over d\cos\theta} d\cos\theta
-\int_{-1}^{0} {d\sigma\over d\cos\theta} d\cos\theta,
\label{generalbf}
\end{equation}
with respect to a CP-odd angular variable $\theta$
. This argument is applicable
for unpolarized or transversely polarized beams as well.

\vskip 0.2cm
\noindent
\underline{\it CP-conjugate process:}
${\cal M}_{--}$ and ${\cal M}_{++}$ are CP
conjugate to each other. 
In this case, instead of a kinematical variable, the
appropriate means to examine CP violation is to 
directly compare the rates of the conjugate processes.
We can thus define another CP asymmetry in 
{\it total cross section rates} 
between the two conjugate processes of opposite helicities,
called the ``left-right'' asymmetry
\begin{equation}
{\cal A}_{LR}=\sigma_{--}-\sigma_{++}.
\label{lr}
\end{equation}

The longitudinally polarized cross section for arbitrary beam
polarizations can be calculated by the helicity amplitudes
\begin{eqnarray}
\nonumber
d\sigma(P_-P_+)={1\over 4}[&&(1+P_-)(1+P_+)d\sigma_{++}+
   (1+P_-)(1-P_+)d\sigma_{+-}\\
+&&(1-P_-)(1+P_+)d\sigma_{-+} 
 + (1-P_-)(1-P_+)d\sigma_{--} ],
\end{eqnarray}
where $P_-\ (P_+)$ is the electron (positron) longitudinal
polarization, with $P_\pm=-1\ (+1)$ for purely left (right)
handed. Whenever appropriate in our later studies, we will
assume the realistic beam polarization as 
$(|P_-|, |P_+|)=(80\%, 60\%)$ \cite{lc}.

\section{CP-odd variables and the CP-odd coupling}

In this section, we construct CP-odd variables for the Higgs
signal in Eq.~(\ref{eeh}) in order to study the CP-violating
interactions in Eq.~(\ref{vert}). Different CP asymmetries
appear to be complementary in exploring different aspects
of the CP-odd coupling $\tilde b$.

\subsection{Simple polar angles and $\Im(\tilde b)$}

\begin{figure}[tb]
%\centerline{\psfig{file=/u/than/temp/fig3_theta.ps,width=3.8in}}
\centerline{\psfig{file=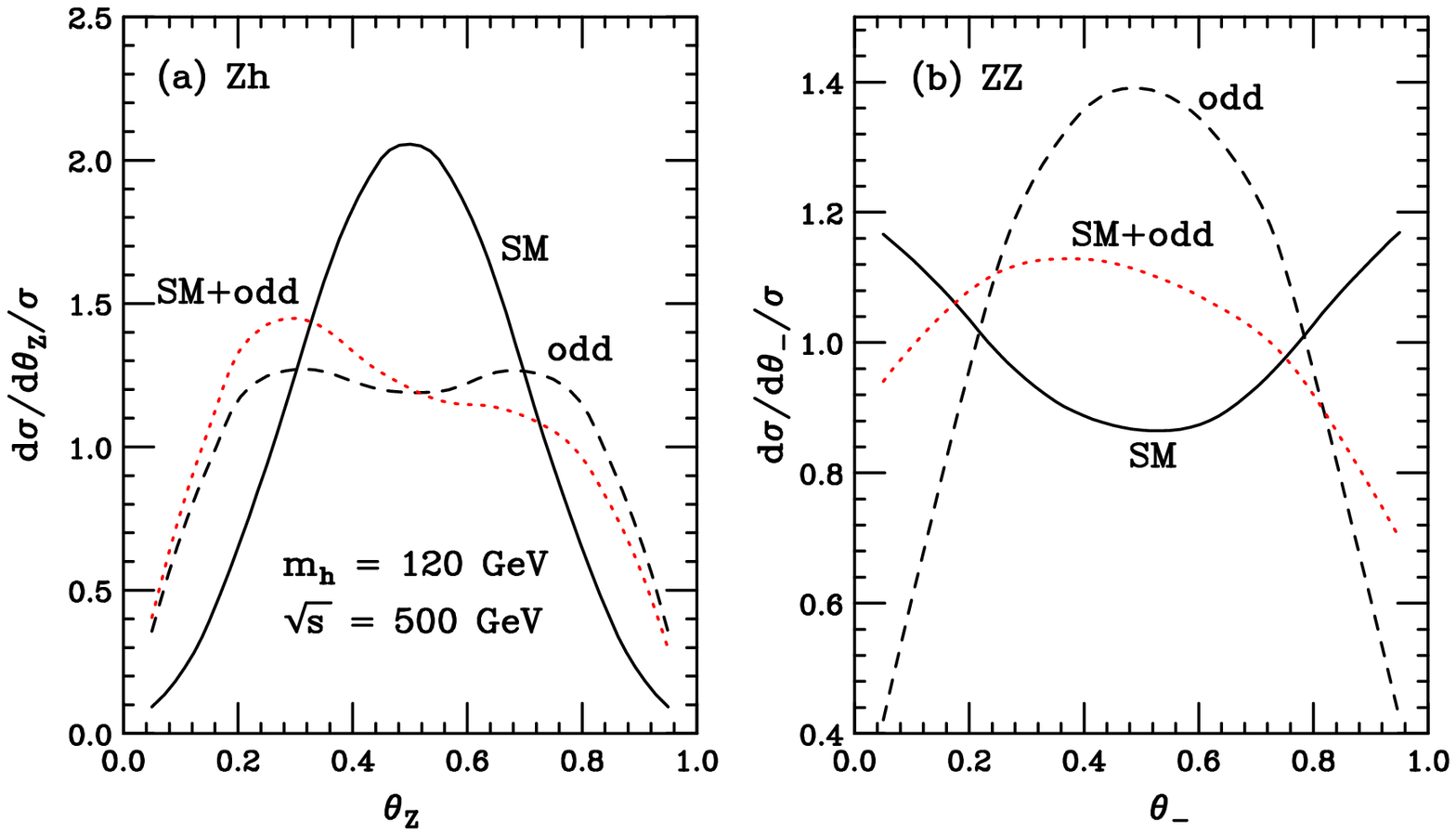,width=5.4in}}
\bigskip
\caption{Normalized polar angle distributions for $\sigma_{-+}$
at $\sqrt s=500\ \gev$ with $\mh=120\ \gev$ for
(a) $\epem \to Z h$ with $Z\to f\bar f$, and
(b) $\epem \to \epem h$ via $ZZ$ fusion. 
The solid curves are for the SM interaction ($a=1$),
the dashed for the CP-odd ($\Im(\tilde b)=1$),
and the dotted for CP violation with $a=\Im(\tilde b)=1$.
$100\%$ longitudinal polarization of $e^-_L e^+_R$ has been used.}
\label{thetaZ}
\end{figure}

It has been argued that the $Zh$ process 
will test the spin-parity property \cite{parity}
of the coupling by simply measuring the polar
angle distribution of the outgoing $Z$ boson.
The distribution can be written in the form
\[ {d\sigma\over d\cos\theta_Z} 
\propto \left\{ \begin{array}{ll}
           \beta^2\sin^2\theta_Z + {8M_Z^2\over s}\quad &\mbox{scalar $h$}\\
           1-{1\over 2}\sin^2\theta_Z\quad &\mbox{pseudo-scalar $h$}
          \end{array}
  \right. \]
In fact, this simple polar angle may provide CP 
information as well.
If we rewrite this angle in terms of a dot-product
\begin{equation}
\cos\theta_Z={\vec p_1\cdot \vec q_+ \over |\vec p_1|\ |\vec q_+|},
\label{thetaz}
\end{equation}
where $\vec q_+=\vec q_1+\vec q_2$ is the vector sum of the outgoing
fermion momenta, it is easy to see that it is
P-odd and C-even under transformation for the final state.
One could thus expect to test CP property of the interactions
by examining the polar angle distribution. The experimental 
study is made particularly simple since this variable 
does not require charge
identification for the final state fermions. Because of this,
one expects to increase the statistical accuracy by including
some well-measured hadronic decay modes of $Z$, as we accept
the light quark jets of Eq.~(\ref{fermions}). However, after 
the azimuthal angle integration the dispersive part of the
form factor proportional to $\Re(\tilde b$) vanishes 
and the surviving term is the absorptive part proportional
to $\Im(\tilde b$).
The angular distributions are shown in Fig.~\ref{thetaZ}(a)
for $\sqrt s=500\ \gev$ with $\mh=120\ \gev$.
The solid curve is for the SM interaction only ($a=1$),
the dashed curve is for the CP-odd only ($\Im(\tilde b)=1$),
the dotted is for CP violation with $a=\Im(\tilde b)=1$.
We see from the dotted curve that there is indeed an asymmetry
with respect to the forward ($\pi/2 \le \theta_Z\le 0$) and 
backward ($\pi\le \theta_Z\le \pi/2$) regions.
We have assumed $100\%$ longitudinal polarization of $e^-_L e^+_R$ 
for illustration here.

\begin{figure}[tb]
%\centerline{\psfig{file=/u/than/temp/fig11_bf_bts.ps,width=5.2in}}
\centerline{\psfig{file=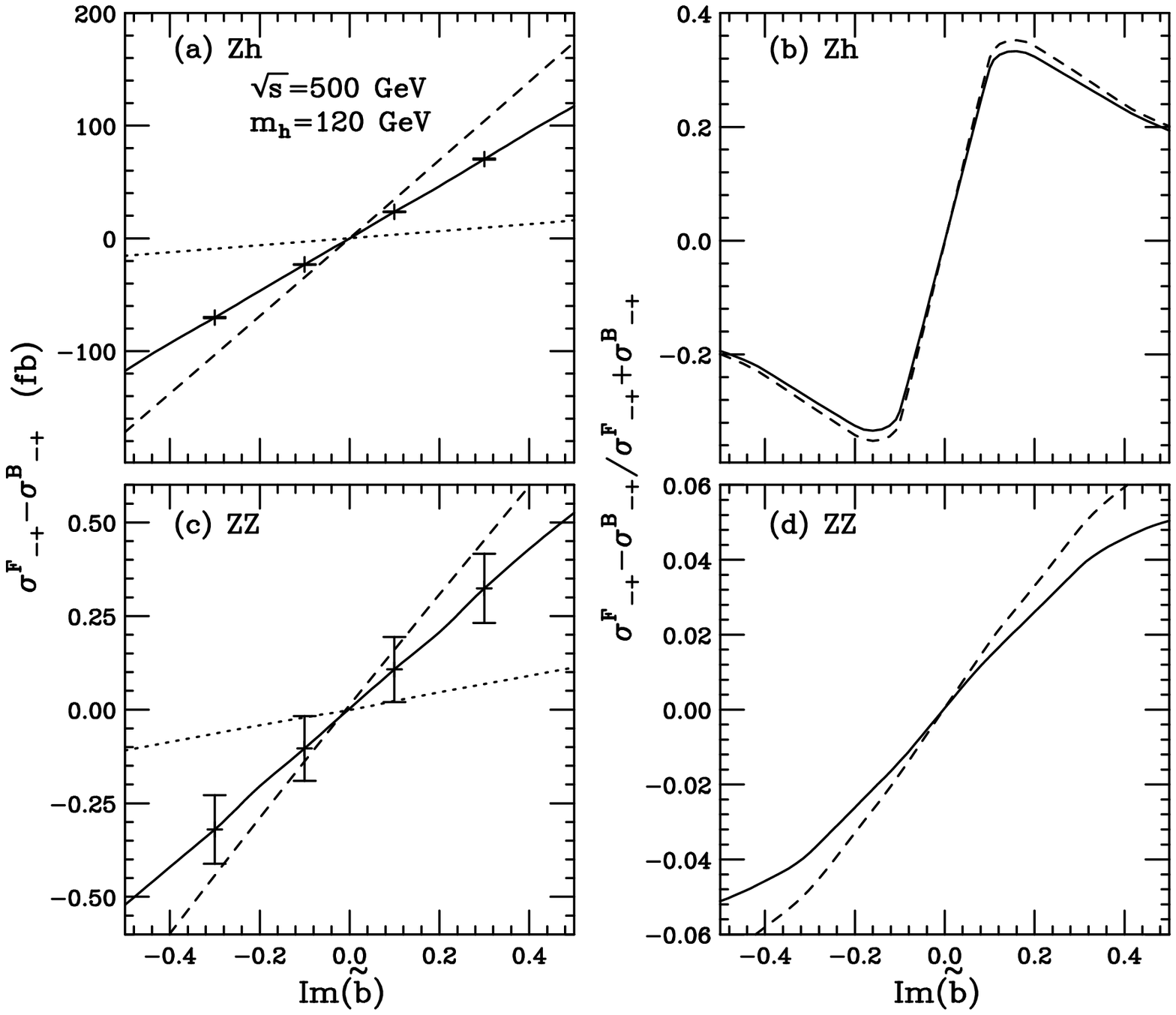,width=5.4in}}
\bigskip
\caption{Forward-backward asymmetries for $\sigma_{-+}$ versus 
$\Im(\tilde b)$ at $\sqrt s=500\ \gev$ with $\mh=120\ \gev$
for (a) $Zh\to f\bar f h$: asymmetry in fb, 
(b) $Zh\to f\bar f h$: percentage asymmetry,
(c) $ZZ$ fusion: asymmetry in fb, 
and (d) $ZZ$ fusion: percentage asymmetry.
The dashed curves are for $100\%$ 
longitudinal polarization $e_L^-e_R^+$, the solid for a realistic 
polarization $(e_L^-,e_R^+)=(80\%,60\%)$, and the dotted for
unpolarized beams. The error bars are 
statistical uncertainties obtained with a luminosity of $1000\ \fbi$.}
\label{bt}
\end{figure}

The above calculation can in principle be carried through for 
the $ZZ$ fusion process. However, due to the unique kinematics
in this process,
it appears that we can define an alternative polar angle 
\begin{equation}
\cos\theta_-={(\vec p_1\times \vec q_-)\cdot (\vec q_1\times \vec q_2)
\over |\vec p_1\times \vec q_-|\ |\vec q_1\times \vec q_2|},
\label{thetam}
\end{equation}
where $\vec q_-=\vec q_1-\vec q_2$, which yields a larger
asymmetry and thus being more sensitive to the coefficient 
$\Im(\tilde b$). It is easy to verify that this variable is P-odd 
and C-even under transformation for the final state. 
Figure~\ref{thetaZ}(b) shows the angular distributions
for $\epem \to \epem h$ via $ZZ$ fusion with 
$100\%$ longitudinal polarization of $e^-_L e^+_R$.
The legend is the same as in (a). 
We see from the dotted curve that an asymmetry exists
with respect to this angle.

\begin{figure}[tb]
%\centerline{\psfig{file=/u/than/temp/fig12_bts_mp.ps,width=5.2in}}
\centerline{\psfig{file=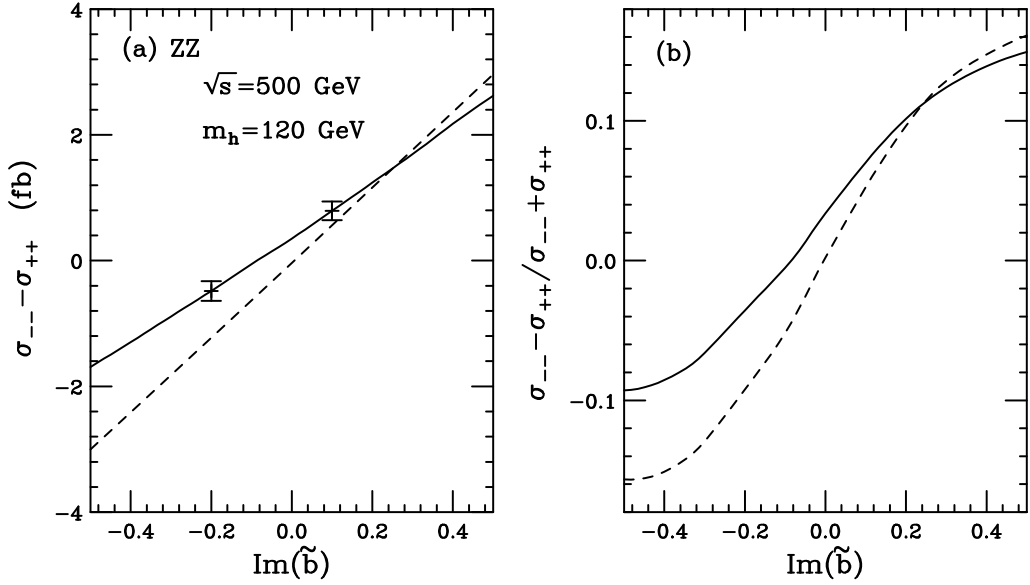,width=5.4in}}
%\centerline{\psfig{file=fig12_bts_mp.ps,width=5.2in}}
\bigskip
\caption{Polarized cross section asymmetry at $\sqrt s=500\ \gev$ 
with $\mh=120\ \gev$ versus $\Im(\tilde b)$ for
(a) the asymmetry in fb, and
(b) the percentage asymmetry. The dashed curves are for $100\%$ 
beam polarization, the solid for a realistic polarization
$(e^-,e^+)=(80\%,60\%)$. The error bars are  
statistical uncertainties obtained with a luminosity of $1000\ \fbi$.} 
\label{bt2}
\end{figure}

Replacing $\theta$ by $\theta_Z$ in Eq.~(\ref{generalbf}),
we can define a forward-backward asymmetry ${\cal A}^{FB}_{\theta_Z}$
with respect to the angle $\theta_Z$,
and similarly another asymmetry ${\cal A}^{FB}_{\theta_-}$ 
with respect to the angle $\theta_-$.
These two asymmetries are calculated for $\sigma_{-+}$, and
shown in Fig.~\ref{bt} at $\sqrt s=500\ \gev$ with 
$\mh=120\ \gev$ versus $\Im(\tilde b)$.
Figures~\ref{bt}(a) and (b) are the asymmetry in fb and 
the percentage asymmetry respectively, with respect to 
$\theta_Z$ in $Zh\to f\bar f h$.
Similarly, Figs.~\ref{bt}(c) and (d) show the asymmetry and
percent asymmetry for $ZZ$ fusion
with respect to $\theta_-$. The dashed curves are for $100\%$
longitudinal polarization $e_L^-e_R^+$, the solid are for a realistic 
polarization $(e_L^-,e_R^+)=(80\%,60\%)$, and the dotted are for
unpolarized beams. We see that the beam polarization here
substantially enhances the asymmetries, and the realistic
polarization maintains the asymmetries to a large extent.
Some degree of asymmetry still exists even for unpolarized
beams. The percentage asymmetry for
the $Zh$ process can be as large as $30\%$ for $\Im(\tilde b)\sim 0.2$,
and is typically of a few percent for $ZZ$ fusion.

\begin{figure}[tb]
%\centerline{\psfig{file=/u/than/temp/fig13_rts.ps,width=5.2in}}
\centerline{\psfig{file=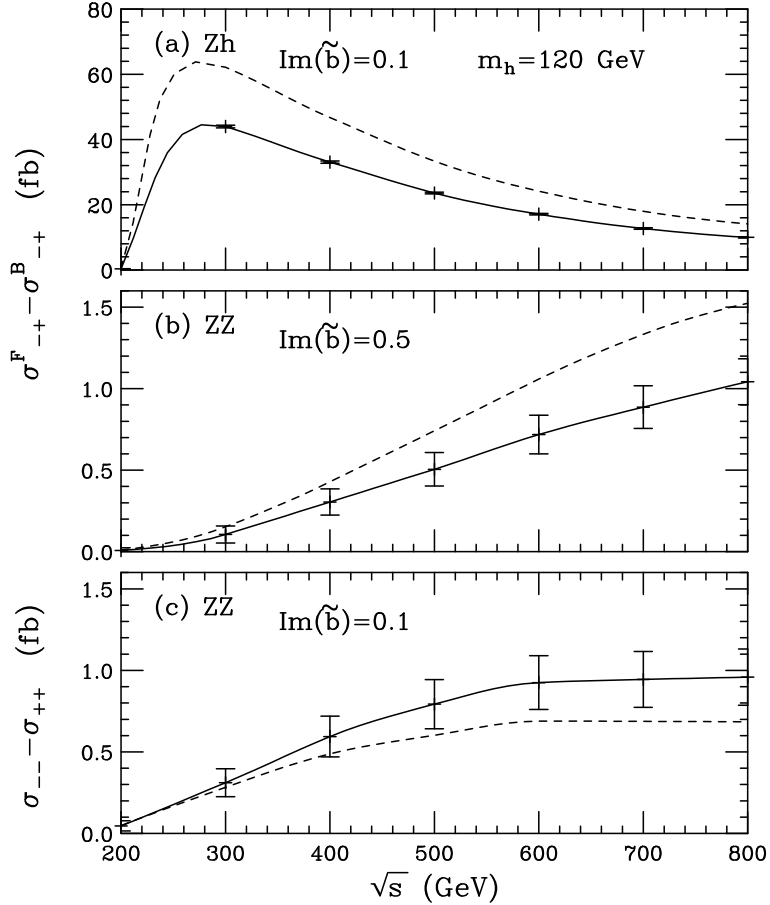,width=4.0in}}
\bigskip
\caption{Cross section asymmetries in fb versus $\sqrt s$ with 
$\mh=120\ \gev$ for
(a) forward-backward asymmetry with respect to $\theta_Z$ 
for $\Im(\tilde b)=0.1$,
(b) forward-backward asymmetry with respect to $\theta_-$ 
for for $\Im(\tilde b)=0.5$,
(c) $LR$ asymmetry between $\sigma_{--}$ and $\sigma-{++}$
for $\Im(\tilde b)=0.1$.
The dashed curves are for $100\%$ longitudinal
polarization, and the solid for a realistic polarization
$(e^-,e^+)=(80\%,60\%)$. The error bars are 
statistical uncertainties obtained with a luminosity of $1000\ \fbi$.}
\label{rts}
\end{figure}

We wish to address to what extent an asymmetry can be determined 
by experiments. For this purpose, we estimate the statistical
uncertainties for the asymmetry measurements.
We determine the Gaussian
statistical error by $\sqrt{N_F+N_B}$ where $N_F\ (N_B)$ 
is the number of forward (backward) events. The statistical 
significance for the asymmetry measurement is obtained by
\begin{equation}
{|N_F - N_B|\over \sqrt{N_F+N_B}}.
\label{error}
\end{equation}
The error bars in the plots are calculted with an assumed integrated 
luminosity of $1000\ \fbi$. Due to the larger asymmetry as well
as a larger cross section for $Zh\to f\bar f h$, the $Zh$
production would provide a much better determination of $\Im(\tilde b$).

As we discussed earlier, the $ZZ$ fusion process can provide
another type of asymmetry between CP conjugate processes, 
in particular between $\sigma_{--}$ and $\sigma_{++}$ as 
defined in Eq.~(\ref{lr}), which is absent in $Zh$ production.
This is presented in Fig.~\ref{bt2} for ${\cal A}_{LR}$,
at $\sqrt s=500\ \gev$ with $\mh=120\ \gev$
versus $\Im(\tilde b)$.
Figure~\ref{bt2}(a) is the asymmetry in fb.
The legend is the same as in Fig.~\ref{bt}. The error bars
are for a total integrated luminosity of $1000\ \fbi$
($500\ \fbi$ each for $\sigma_{--}$ and $\sigma_{++}$).
The percentage asymmetry in Fig.~\ref{bt2}(b) can be
at a $10\%$ level for $\Im(\tilde b)\sim 0.2$.
It is interesting to note that the solid curves yield
a non-zero value for $\tilde b=0$. This is due to the
intrinsic $LR$ asymmetry of the $Z$ coupling to electrons.
This shift appears when $|P_-|\ne |P_+|$ and is proportional 
to $\sigma_{-+}-\sigma_{+-}$. It can be well predicted in the 
SM for a given beam polarization.

\begin{figure}[tb]
%\centerline{\psfig{file=/u/than/temp/fig3_theta.ps,width=3.8in}}
\centerline{\psfig{file=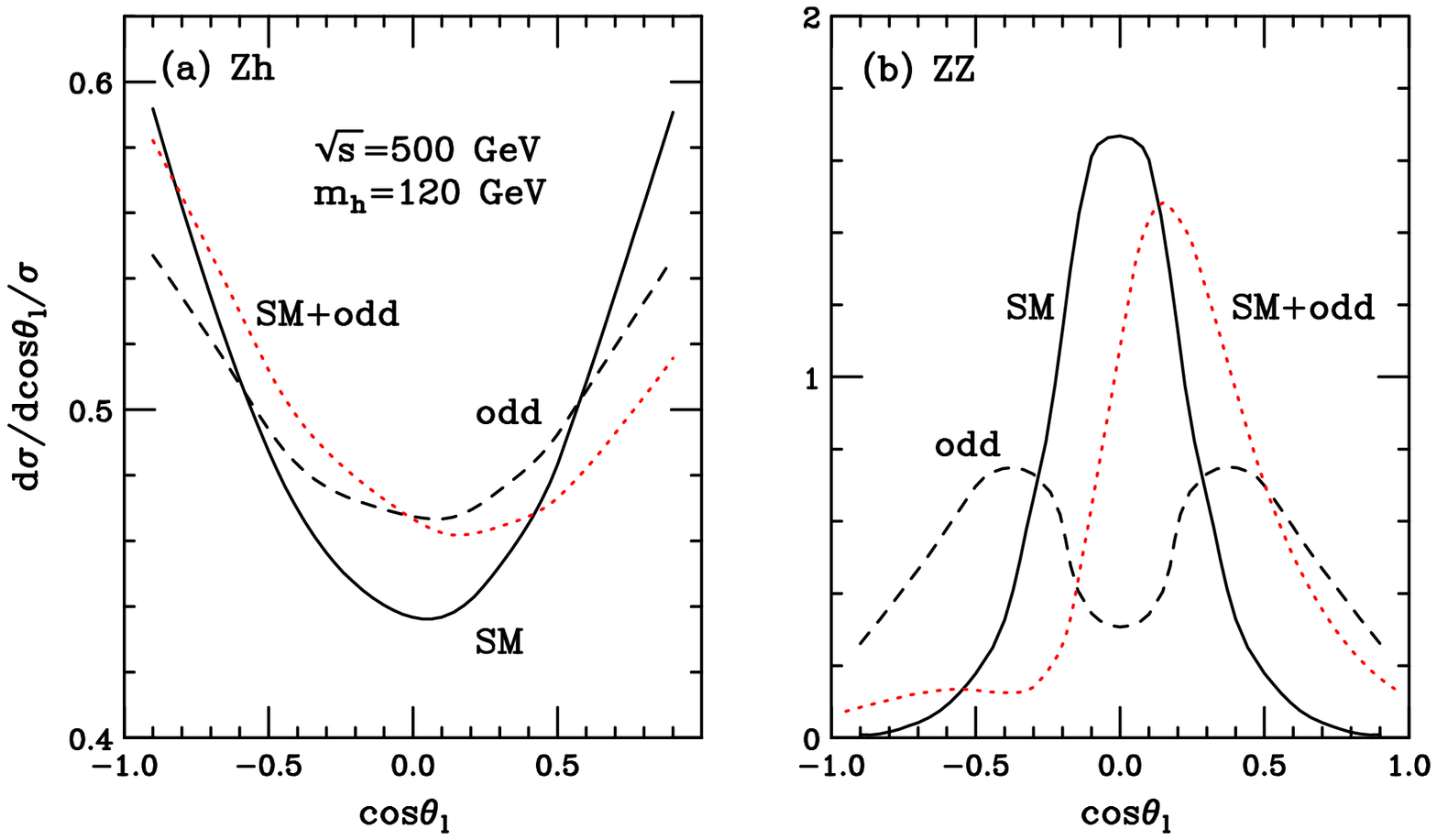,width=5.4in}}
\bigskip
\caption{Normalized angular distributions for $\sigma_{-+}$
at $\sqrt s=500\ \gev$ with $\mh=120\ \gev$ for
(a) $\epem \to Z h$ with $Z\to e^-e^+,\mu^-\mu^+$, and
(b) $\epem \to \epem h$ via $ZZ$ fusion. 
The solid curves are for the SM interaction ($a=1$),
the dashed for the CP-odd ($\Re(\tilde b)=1$), and
the dotted for CP violation with $a=\Re(\tilde b)=1$.
$100\%$ longitudinal polarization of $e^-_L e^+_R$ has been used.}
\label{thetal}
\end{figure}

Cross section asymmetries versus $\sqrt s$
are shown in Fig.~\ref{rts} in units of fb with $\mh=120\ \gev$
(a) forward-backward asymmetry for $\sigma_{-+}$ in $Zh$
production with respect to $\theta_Z$ for $\Im(\tilde b)=0.1$,
(b) forward-backward asymmetry for $\sigma_{-+}$ in $ZZ$ fusion
with respect to $\theta_-$ for for $\Im(\tilde b)=0.5$,
(c) $LR$ asymmetry between $\sigma_{--}$ and $\sigma_{++}$
for $\Im(\tilde b)=0.1$.
The dashed curves are for $100\%$ longitudinal polarization, 
and the solid are for a realistic polarization
$(e^-,e^+)=(80\%,60\%)$. The error bars are for the 
statistical uncertainty with a luminosity of $1000\ \fbi$.
We see again the possibly good accuracy for determining the
asymmetry by the $Zh$ process. Furthermore, these two processes
are complementary: at lower energies near threshold the $Zh$
production is far more important; while at higher energies
the $ZZ$ fusion becomes increasingly significant, as has been
seen in Fig.~\ref{tot}. In Fig.~\ref{rts}(c), the reason
that the realistic asymmetry (solid) is even bigger than
the ideal case (dashed) is due to the non-zero contribution
from the CP-conserving $LR$ asymmetry of the $Z$ coupling
as discussed in the last paragraph.

\subsection{The lepton momentum orientation and $\Re(\tilde b)$}

\begin{figure}[tb]
%\centerline{\psfig{file=/u/than/temp/fig11_bf_bts.ps,width=5.2in}}
\centerline{\psfig{file=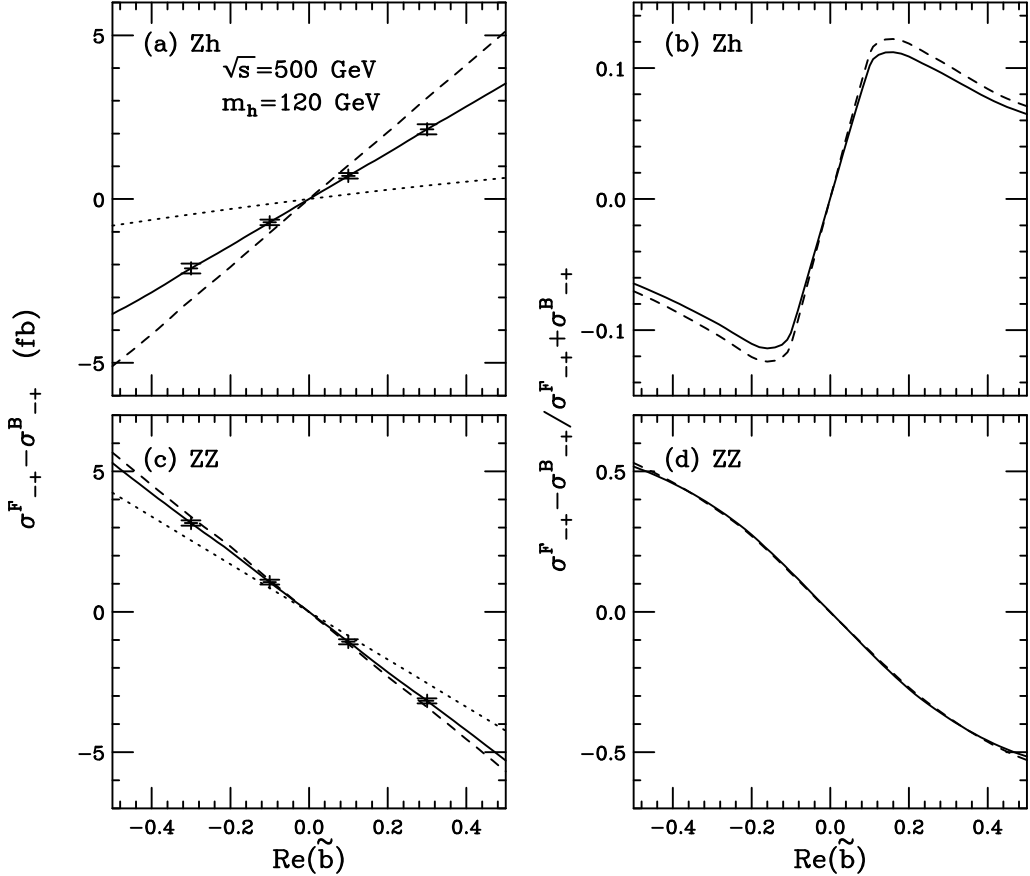,width=5.4in}}
\bigskip
\caption{Forward-backward asymmetry for $\sigma_{-+}$ versus 
$\Re(\tilde b)$ at $\sqrt s=500\ \gev$ with $\mh=120\ \gev$
for (a) $Zh\to f\bar f h$: asymmetry in fb, 
(b) $Zh\to f\bar f h$: percentage asymmetry,
(c) $ZZ$ fusion: asymmetry in fb, 
and (d) $ZZ$ fusion: percentage asymmetry.
The dashed curves are for $100\%$ 
longitudinal polarization $e_L^-e_R^+$, the solid for a realistic 
polarization $(e_L^-,e_R^+)=(80\%,60\%)$, and the dotted for
unpolarized beams. The error bars are 
statistical uncertainties obtained with a luminosity of $1000\ \fbi$.}
\label{btl}
\end{figure}

We showed in the last section that the simple polar angles
can probe CP violation for a Higgs-gauge boson coupling,
but only for the absorptive part of the form factor 
$\Im(\tilde b$). In order to be sensitive to the dispersive
part $\Re(\tilde b$), one needs to construct more sophisticated
variables, involving the azimuthal angle information for the
final state fermions. We find that a simple variable to
serve this purpose \cite{germanagain} can be defined as
\begin{equation}
\cos\theta_\ell={\vec p_1\cdot (\vec q_1\times \vec q_2)
\over |\vec p_1|\ |\vec q_1\times \vec q_2|},
\label{theta0}
\end{equation}
where $\vec q_1\times \vec q_2$ defines the orientation of
the plane for the final state fermion pair. This variable
is P-even and C-odd under final state transformation. However,
we would need to unambiguously identify the fermion from
the anti-fermion, and to accurately determine their
momenta. This is naturally achievable for the
$ZZ$ fusion process, while we will have to limit ourself
to $f=e^-,\mu^-$ for the $Zh\to f\bar f h$ process.
Explicit calculations show that this variable is
only sensitive to $\Re(\tilde b)$ and insensitive
to $\Im(\tilde b)$. 

We evaluate the angular distribution for $\cos\theta_\ell$ 
at $\sqrt s=500\ \gev$ with $\mh=120\ \gev$. Shown in
Fig.~\ref{thetal} are the normalized distributions for
(a) $\epem \to Z h$ with $Z\to e^-e^+,\mu^-\mu^+$, 
and (b) $\epem \to \epem h$ via $ZZ$ fusion. 
The solid curves are for the SM interaction ($a=1$),
the dashed curves are for the CP-odd ($\Re(\tilde b)=1$),
the dotted are for CP violation with $a=\Re(\tilde b)=1$.
$100\%$ longitudinal polarization of $e^-_L e^+_R$ has been used
as for $\sigma_{-+}$. The CP asymmetries are manifest
as seen from the dotted curves.
We define a CP asymmetry ${\cal A}^{FB}_{\theta_\ell}$ 
in the same way as in Eq.~(\ref{generalbf}).
The asymmetries for these two processes
are calculated for $\sigma_{-+}$, and
shown in Fig.~\ref{btl} at $\sqrt s=500\ \gev$ with 
$\mh=120\ \gev$ versus $\Re(\tilde b)$. The parameters
and legend are the same as in Fig.~\ref{bt}.
We see that the percentage asymmetry for
the $Zh$ process is about $10\%$ percent and for
$ZZ$ fusion it can be as large as $30\%$ for 
$\Re(\tilde b)\approx 0.2$. 
The error bars in the plots are estimated with an integrated 
luminosity of $1000\ \fbi$. Due to the large asymmetries
both $Zh$ production and $ZZ$ fusion processes could 
provide a good probe to the coupling $\Re(\tilde b$).
A particularly important result as indicated in
Figs.~\ref{btl}(c) and (d) is that the
asymmetry for the $ZZ$ fusion is rather insensitive
to the beam polarization.

\begin{figure}[tb]
%\centerline{\psfig{file=/u/than/temp/fig13_rts.ps,width=5.2in}}
\centerline{\psfig{file=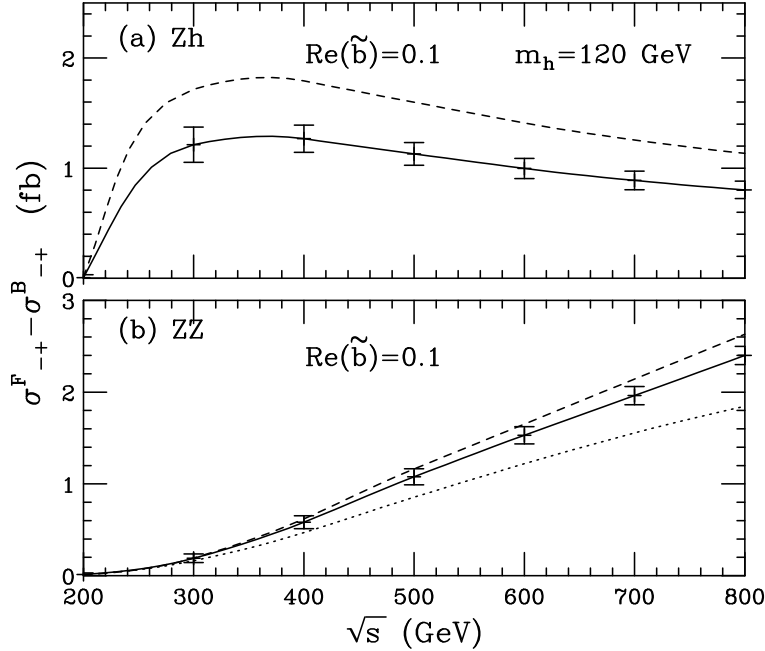,width=4.0in}}
\bigskip
\caption{Forward-backward cross section asymmetries for $\sigma_{-+}$ 
with respect to $\theta_\ell$ in fb 
versus $\sqrt s$ with $\mh=120\ \gev$ and $\Re(\tilde b)=0.1$ for
(a) $Zh$ production and (b) $ZZ$ fusion.
The dashed curves are for $100\%$ longitudinal 
polarization, and the solid for a realistic polarization
$(e^-,e^+)=(80\%,60\%)$. The error bars are  
statistical uncertainties obtained with a luminosity of $1000\ \fbi$.}
\label{rtsl}
\end{figure}

Forward-backward cross section asymmetries for $\sigma_{-+}$ 
with respect to $\theta_\ell$
are shown versus $\sqrt s$ in Fig.~\ref{rtsl} 
with $\mh=120\ \gev$ and $\Re(\tilde b)=0.1$. 
Figure~\ref{rtsl}(a) is the asymmetry for $Zh$
production, and (b) is for $ZZ$ fusion. 
We see again good sensitivity for measuring the
asymmetry especially by the $ZZ$ fusion process
and at higher energies, which appears to have very
little dependence on the beam polarization.

To further assess the linear collider sensitivity 
to $\tilde b$, we compare all the CP asymmetries and
present in Table~\ref{tab} the $95\%$ Confidence 
Level ($2\sigma$) sensitivity limits with $\mh=120\ \gev$
for two collider energies $\sqrt s=500,\ 800\ \gev$ 
and two choices of integrated luminosity
${\cal L}=500,\ 1000\ {\rm\ fb^{-1}}$.
Realistic polarizations of $(80\%, 60\%)$ are used unless 
specified for no beam polarization by ``unpolarized''.
We see that at a 500 GeV linear
collider with a total luminosity of $1000\ \fbi$, the 
CP-odd coupling form factor may be sensitively probed
to a value of about $\Im(\tilde b)\approx 0.0022$
and $\Re(\tilde b)\approx 0.017$ at a $95\%$ C.L.
The coupling may even be probed without a beam polarization
to a level of about 
$\Im(\tilde b)\approx 0.013$ and $\Re(\tilde b)\approx 0.018$. 
The beam polarization improves the sensitivity to
$\Im(\tilde b)$ by about a factor of $5-6$ 
via ${\cal A}^{FB}_{\theta_Z}(Zh)$, but does little to
$\Re(\tilde b)$ through ${\cal A}^{FB}_{\theta_\ell}(ZZ)$.
At $\sqrt s=800\ \gev$, the sensitivity in $Zh$ process is
slightly degraded. On the other hand, the sensitivity
in $ZZ$ fusion is enhanced by about a factor of two
due to the larger cross section and larger asymmetry
at higher energies.

\begin{table}[tb]
\begin{tabular}{llcccc}
%& & \multicolumn{3}{c}{${\cal L}\rm\ (fb^{-1})$}\\
%\cline{3-5}
&$\sqrt s\ (\gev)$ & 500& 500& 800& 800\\
&${\cal L}\rm\ (fb^{-1})$ & 500& 1000 & 500& 1000\\
\hline\hline
               & ${\cal A}^{FB}_{\theta_Z}(Zh)\ [-+]$& 
0.0028& 0.0022& 0.0043& 0.0032\\
$\Im(\tilde b)$& ${\cal A}^{FB}_{\theta_Z}(Zh)$\ [unpol.]& 
0.019& 0.013& 0.025& 0.019\\
\cline{2-6}
               & ${\cal A}^{FB}_{\theta_-}(ZZ)\ [-+]$ & 
0.21& 0.16& 0.19& 0.13\\ 
\cline{2-6}
               & ${\cal A}_{LR} (ZZ)$& 0.071& 0.045& 0.065& 0.041\\
\hline\hline
               & ${\cal A}^{FB}_{\theta_\ell}(Zh)\ [-+]$
& 0.023& 0.018& 0.019& 0.014\\
\cline{2-6}
$\Re(\tilde b)$& ${\cal A}^{FB}_{\theta_\ell}(ZZ)\ [-+]$
& 0.021& 0.017& 0.014& 0.009\\
               & ${\cal A}^{FB}_{\theta_\ell}(ZZ)$\ [unpol.]& 
0.024& 0.018& 0.016& 0.010
\end{tabular}
\caption{$95\%$ C.L. limits on $\tilde b$ from the CP asymmetries
defined in the text at $\sqrt s=500,\ 800\ \gev$ with $\mh=120\ \gev$,
for two representative luminosities
${\cal L}=500,\ 1000\ {\rm\ fb^{-1}}$.
Realistic polarizations
of $(80\%, 60\%)$ are used unless specified as ``unpolarized''.}
\label{tab}
\end{table}

\section{Discussions and Conclusions}

Before summarizing our results, 
a few remarks are in order. First, in 
previous studies of the $Zh$ process \cite{parity,eezh},
a common variable is defined as
\begin{equation}
\cos\theta_+={(\vec p_1\times \vec q_+)\cdot (\vec q_1\times \vec q_2)
\over |\vec p_1\times \vec q_-|\ |\vec q_1\times \vec q_2|},
\label{thetap}
\end{equation}
where $\vec q_+=\vec q_1+\vec q_2=\vec p_Z$. This variable
seems quite suitable for the $Zh$ production since it is the
azimuthal angle formed between the $Zh$ production plane and
the decay plane of $Z\to f\bar f$
if the $Z$ momentum is chosen to define the
rotational axis.
However, this variable is P-even and C-even under
final state transformation and thus cannot provide
an unambiguous measure for CP violation alone. One would
have to analyze other angular distributions to extract
the CP property of the interaction.

As a second remark, one may consider our analysis
for the $ZZ$ fusion similar to that in $e^-e^-$ 
collisions \cite{emem},
since the only tree-level Higgs boson production at
$e^-e^-$ colliders is via the $ZZ$ fusion 
mechanism \cite{hanemem}. 
However, an $e^-e^-$ initial state cannot be made a CP eigenstate
as evident from the discussion of Eq.~(\ref{cp}). 
The explicit CP asymmetry in $e^-e^-$ collisions would have 
to be constructed in comparison with the conjugate 
$e^+e^+$ reactions.

Finally, although the $ZZh$ coupling under current investigation
is arguably the most important interaction in the light of 
electroweak symmetry breaking, other interaction vertices 
such as $Z\gamma h$ and $\gamma\gamma h$ may be equally 
possible to contain CP violation induced by loop effects.
Although the CP asymmetries constructed in this paper should
be generically applicable to the other cases as well,
we choose not to include those coupling in our analyses
for the sake of simplicity. However, in terms of our
$ZZ$ fusion study, since the photon-induced processes 
$\gamma\gamma \to h, \gamma Z\to h$ would mainly give 
collinear electrons along the beams, 
our kinematical requirement to tag $\epem$ final 
state at a large angle 
will effectively single out the $ZZh$ contribution.

To summarize our analyses of possible CP violation 
for the interaction vertex $ZZh$, 
we classified the signal channel into two categories
as $Zh$ production with $Z\to f\bar f$ and $ZZ$ fusion. 
We proposed four simple CP-asymmetric variables 
\begin{eqnarray}
\nonumber
&&{\cal A}^{FB}_{\theta_Z}:\qquad {\rm for}\ Zh\ {\rm production},\\
\nonumber
&&{\cal A}^{FB}_{\theta_-}:\qquad {\rm for}\ ZZ\ {\rm fusion},\\
\nonumber
&&{\cal A}_{LR}:\qquad {\rm for}\ ZZ\ {\rm fusion\ only},\\
\nonumber
&&{\cal A}^{FB}_{\theta_\ell}:\qquad {\rm for\ both}\ Zh,\ ZZ.
\end{eqnarray}
We found them complementary in probing the CP-odd coupling form factor
$\tilde b$. The first three are sensitive to $\Im(\tilde b)$,
while the last one sensitive to $\Re(\tilde b)$. 
${\cal A}^{FB}_{\theta_Z}$ yields the largest asymmetry
for $\Im(\tilde b)$ (see Fig.~\ref{bt}), 
while ${\cal A}^{FB}_{\theta_\ell}$
is the largest for $\Re(\tilde b)$ (see Fig.~\ref{btl}), 
both reaching about
$30\%$ for $|\tilde b|\approx 0.2$. 
The ultimate sensitivity to $\tilde b$ depends on both the
size of asymmetry and the signal production rate.
As illustrated in Table~\ref{tab}, at a 500 GeV linear
collider with a total luminosity of $1000\ \fbi$, the 
CP-odd coupling may be sensitively probed
to a value of about $\Im(\tilde b)\approx 0.0022$
and $\Re(\tilde b)\approx 0.017$ at a $95\%$ C.L.
with the beam polarization $(80\%, 60\%)$. The coupling
may even be probed without beam polarization to a level of about 
$\Im(\tilde b)\approx 0.013$ and $\Re(\tilde b)\approx 0.018$. 
At a higher energy collider with
$\sqrt s=800\ \gev$, the sensitivity in $Zh$ process is
slightly degraded but that in $ZZ$ fusion is enhanced by
about a factor of two.

\vskip 0.2cm
{\it Acknowledgments}:
We thank R. Sobey for his early participation
of this project. We would also like to thank K. Hagiwara,
W.-Y. Keung, G. Valencia and P. Zerwas for helpful discussions.  
This work was supported in part by
a DOE grant No. DE-FG02-95ER40896 and in part by 
the Wisconsin Alumni Research Foundation.

\end{document}